\documentclass[12pt]{amsart}
\usepackage{amsmath}  

\newtheorem{theorem}{Theorem}
\newtheorem{proposition}{Proposition}

\def\<{\langle}
\def\>{\rangle}
\def\d{{\mathrm d}}
\def\bo{\mathcal B([0,2\pi))}

\begin{document}

\sloppy

\title{The Pegg-Barnett formalism and covariant phase observables}

\author{Pekka Lahti}\address{Department of Physics, 
University of Turku, 20014 Turku, Finland}\email{pekka.lahti@utu.fi}
\author{Juha-Pekka Pellonp\"a\"a}\address{Department of Physics, 
University of Turku, 20014 Turku, Finland}\email{juhpello@utu.fi}
\maketitle
\begin{abstract}
We compare the Pegg-Barnett (PB) formalism with the covariant phase observable 
approach to the problem of quantum phase 
and show that PB-formalism gives essentially 
the same results as the canonical (covariant) phase observable.
We also show that PB-formalism can be extended to cover all covariant
phase observables including the covariant 
phase observable arising from the angle 
margin of the Husimi $Q$-function.
\end{abstract}

\section{Introduction}
There are at least three different approaches to the problem of quantum phase.
First, one may quantize some appropriate classical dynamical variable, e.g.\
the phase angle of two dimensional phase space, using some quantization rule
to get a self-adjoint phase operator.
The most widely used quantization rule is the Cahill-Glauber $s$-parametrized \cite{CaGl} 
quantization and especially one of its special cases, 
the Wigner-Weyl quantization (see, e.g.\ \cite{Ro, DuHe}).
Second, in the Pegg-Barnett (PB) formalism, one may calculate
the phase properties of a single mode electromagnetic field using a
sequence of self-adjoint operators acting on finite subspaces of 
the infinite dimensional Hilbert space of the single mode system (see, e.g.\
\cite{PeBa}).
The third way to approach the phase problem is to extend 
the mathematical representation of the concept of 
quantum observable from a self-adjoint operator to a normalized positive operator
measure (POM, for short) and assume that any quantum phase observable 
is a phase shift covariant POM with the interval $[0,2\pi)$ as 
the range of its possible measurement outcomes.

It has been shown \cite{Pe} that the $s$-quantized phase angle operators
arise from phase shift covariant generalized operator measures.
They are covariant POMs when the quantization parameter $s$ is suitably chosen but,
for example, in the case of the Wigner-Weyl quantized phase angle this is not the case
\cite{Pe, DuHe2}.
Still, the structure of the $s$-quantized phase angle operators is similar to
the first moment operators of phase shift covariant POMs.
The PB-formalism can also be embedded in the covariant approach and in this
article we show that the PB-formalism can be extended to cover the
whole covariant theory.
Before doing this, we recall some basic physical properties which are 
important in the theory of quantum phase.

As it appears from the abundant literature, the main
principles used to define a phase observable are the following:
\begin{itemize}
\item[  P0.] The range of values of a phase observable is the interval $[0,2\pi)$.
\item[  P1.] The number observable generates phase shifts.
\item[  P2.]  The phase is completely random in the number states.
\item[  P3.]  A phase observable generates number shifts.
\end{itemize}
Within the POM approach, these postulates can be incorporated in a logical
fashion as follows:
P0 and P1 lead to define a phase observable as a phase shift covariant POM
based on $[0,2\pi)$.  Any such operator measure fulfils P2, but P0 and P2 do not imply P1.
P3 requires that a phase observable generates number shifts, which forces a phase shift covariant
POM to be strong.  Then P0, P1, and P3 imply that the phase observable is 
the canonical (covariant) phase observable  (modulo unitary equivalence).
Thus, the canonical phase observable is (up to unitary equivalence) the only phase observable which
has all the four physically relevant properties P0--P4.

The structure of this article is as follows.
In Sec.\ 2 we give our basic notations and definitions.
In Sec.\ 3 we show that phase shift covariance is
a natural condition for phase observables which describe
phase measurements in coherent states.
In Sec.\ 4 we introduce covariant phase observables and,
finally, in Sec.\ 5 we show the connections between the PB-formalism and
the covariant phase observable approach.

\section{Basic notations}

Let $\mathcal H$ be a complex separable Hilbert space, and  fix an orthonormal basis
$\{|n\rangle\in\mathcal H\,|\,n=0,1,2,...\}$ on it.  We call it the number basis. With respect to
that  we define the lowering operator
$a:=\sum_{n=0}^\infty \sqrt{n+1}|n\rangle\langle n+1|$,
its adjoint the raising operator
$a^*:=\sum_{n=0}^\infty \sqrt{n+1}|n+1\rangle\langle n|$,
and the number operator
$N:=a^*a=\sum_{n=0}^\infty n|n\rangle\langle n|$, with their
usual domains. 
The unitary operators
$R(\theta):=e^{i\theta N}$, $\theta\in\mathbb R$,  are called phase shifters.
If 
 $|z\rangle = e^{-|z|^2/2}\sum_{n=0}^\infty \frac{z^n}{\sqrt{n!}}|n\>$, 
$z\in\mathbb C$, is  a coherent state, then  
$R(\theta)|z\rangle=\left|ze^{i\theta}\right\rangle$ for all $\theta\in\mathbb R$, showing 
the fundamental fact that
the number operator generates phase shifts in coherent states.

Let
 $\mathcal L(\mathcal H)$ denote the set of bounded operators on $\mathcal H$, and
 $\mathcal B(\Omega)$ the $\sigma$-algebra of the Borel subsets of a set
$\Omega\subseteq\mathbb C$.
Let $E:\,\mathcal B(\Omega)\to\mathcal L(\mathcal H)$ be  
a normalized positive operator measure. We recall that  a map 
$E:\,\mathcal B(\Omega)\to\mathcal L(\mathcal H)$ 
is a POM
if and only if  for any unit vector
$\varphi\in\mathcal H$, the map $X\mapsto p^E_\varphi(X) :=
\<\varphi|E(X)\varphi\>$ is a probability measure.
We also recall that a POM $E:\,\mathcal B(\Omega)\to\mathcal L(\mathcal H)$ is
projection valued,  that is,  $E(X)^2=E(X)$ for all $X\in\mathcal B(\Omega)$, if 
and only if it is multiplicative, that is, $E(X\cap Y)=E(X)E(Y)$ for all $X,Y\in\mathcal B(\Omega)$.
%[This is standard piece of math, but some references could be given if wishful].
We say that a POM $E:\,\bo\to\mathcal L(\mathcal H)$ is
{\em phase shift covariant} if 
\begin{equation}\label{yx}
R(\theta)E(X)R(\theta)^*=E(X\oplus\theta)
\end{equation}
for all $X\in\bo$, $\theta\in\mathbb R$,  
where  $X\oplus\theta:=\{x\in[0,2\pi)\,|\,(x-\theta)\mod 2\pi\in X\}$,
that is,
for any unit vector $\varphi\in\mathcal H$
$$
p^E_{R(-\theta)\varphi}(X)=p^E_\varphi(X\oplus\theta)
$$
for all $X\in\bo$, $\theta\in\mathbb R$.
We take Eq.\ (\ref{yx}) to formalize Condition P1.

\section{Coherent states and phase shift covariant operator measures}

Let $|z\rangle$ be a coherent state,
 $E:\,\bo\to\mathcal{L(H)}$  a POM, and
$p^E_{|z\>}$
the corresponding  probability measure.
In order that the numbers $p^E_{|z\>}(X)$ could be interpreted as measurement outcome
probabilities for the coherent state phase measurements,
it is natural to require that they fullfill the following
covariance condition: 
\begin{equation}\label{cov}
p^E_{|ze^{-i\theta}\>}(X)=p^E_{|z\>}(X\oplus\theta),
\end{equation}
for all $z\in\mathbb C$, $\theta\in[0,2\pi)$, and $X\in\bo$,
that is,  
the probability for a phase measurement to give a result in the set $X$ 
in the phase shifted coherent state  $R(-\theta)|z\>$ is equal to 
the probability that the measurement 
in the coherent state $|z\>$ leads to a result in the shifted set $X\oplus\theta$.
The following result is a well-known consequence of the properties
of coherent states (see, e.g.\ \cite[p.\ 13]{Kl}). For completeness
we give here its simple direct proof.

\begin{theorem}
Let $E:\,\bo\to\mathcal{L(H)}$ be a POM. If for all $z\in\mathbb C$,
the coherent state probability measures $p^E_{|z\>}$
satisfy the covariance condition (\ref{cov}), then $E$ is phase shift covariant.
\end{theorem}
\begin{proof}
Let $A\in\mathcal{L(H)}$, and denote its matrix elements with respect to the number
states as   $A_{n,m}=\<n|A|m\>, n,m\in\mathbb N$.  Assume that 
$\<z|A|z\>=0$ for all $z\in\mathbb C$. Since
$$
\<z|A|z\>=e^{-|z|^2}\sum_{n,m=0}^\infty A_{n,m}\frac{|z|^{n+m}}{\sqrt{n!m!}}
e^{i(n-m)\arg z}
$$
is a Fourier series with absolute convergence, it follows that
the Fourier coefficient
$$
f_k(x):=\sum_{n=0}^\infty A_{n,n+k}\frac{x^n}{\sqrt{n!(n+k)!}}=0
$$
for all $x\in[0,\infty)$ and $k\in\mathbb N$.
Due to the uniform convergence,
$$
0=\frac{\d^sf_k}{\d x^s}(0)=A_{s,s+k}\sqrt{\frac{s!}{(s+k)!}}
$$
for all $s\in\mathbb N$, the
matrix elements $A_{n,m}=0$ for all $m\ge n$. Similarly, one proves that
$A_{n,m}=0$ for all $m<n$, so that $A=O$.
Since (\ref{cov}) equals the condition
$$\<z|[E(X\oplus\theta)-R(\theta)E(X)R(\theta)^*]|z\>=0$$ 
for all $z\in\mathbb C$, $X\in\bo$, and $\theta\in\mathbb R$, the theorem follows.
\end{proof} 
We call a phase shift covariant normalized positive operator measure
$E:\,\bo\to\mathcal{L(H)}$ a {\em covariant phase observable}.
Thus, a covariant phase observable is a POM which satisfies Conditions P0 
and P1 (or equivalently P0 and Eq.\ (\ref{cov})).

\section{Covariant phase observables and their operators}
A positive semidefinite complex matrix $(c_{n,m})_{n,m\in\mathbb N}$ with
$c_{n,n}=1$, $n\in\mathbb N$, is called a phase matrix.
It is known that 
$(c_{n,m})_{n,m\in\mathbb N}$ is a phase matrix if and only if
$
c_{n,m}=\langle\psi _n|\psi_m\rangle
$
for some sequence $(\psi_n)_{n\in\mathbb N}\subset\mathcal H$ of unit vectors \cite{BCR,
CDVLP01}.
Then 
$E:\,\mathcal B[0,2\pi)\to\mathcal{L(H)}$ is a covariant phase observable if and only if
$$
E(X)=\sum_{n,m=0}^\infty c_{n,m}\frac{1}{2\pi}\int_Xe^{i(n-m)\theta}\mathrm{d}\theta
|n\rangle\langle m|
$$
where $(c_{n,m})_{n,m\in\mathbb N}$ is a phase matrix, 
$|n\>\<m|$ is the rank-one operator $\mathcal H\ni\psi\mapsto\<m|\psi\>|n\>\in\mathcal H$,
and where the series converges in
the weak operator topology \cite{Holevo, JPPL99, CDVLP01}. 
It is well known that
a covariant phase observable  cannot be projection valued.
Clearly, the support of a covariant phase observable is the closed phase interval
$[0,2\pi]$.

{\em For any covariant phase observable $E$  the phase distribution in the number state $|n\>$ is
uniform}. Indeed, $p^E_{|n\>}(X) = \frac 1{2\pi}\int_X \mathrm{d}\theta$ for all $X\in\bo$.
This is a fundamental property of a phase observable. However, it does  not force a
POM to be covariant under the phase shifts generated by the number observable. In fact,
consider the following  normalized positive operator measure
$$
F:\,\bo\to\mathcal{L(H)},\;X\mapsto \frac 1{2\pi}\int_X \mathrm{d}\theta\,I+
\frac 1{2\pi}\int_X\sin\theta\,\mathrm{d}\theta\,(|0\>\<1|+|1\>\<0|)
$$
Clearly, the number state  probability measures $p^F_{|n\>}$ are uniform,
$\<n|F(X)|n\>=\frac 1{2\pi}\int_X \mathrm{d}\theta$ for all $X\in\bo$, but 
$F$ is not phase shift covariant. This shows that P0 and P2 do not imply P1.
 
For any continuous function $f:\,[0,2\pi]\to\mathbb C$ the integral of $f$ with respect
to $E$ is a (weakly defined) bounded operator denoted as 
$\int_0^{2\pi}f(\theta)\d E(\theta)$. 
In particular, the moment operators of the covariant phase observable $E$,
$$
E^{(k)}:=\int_0^{2\pi}\theta^k\mathrm dE(\theta),\;\;\;k\in\mathbb N,
%\pi I+\sum_{n\ne m=0}^\infty c_{n,m}\frac{i}{m-n}|n\rangle\langle m|
$$
are bounded self-adjoint operators. 
It is a well-known consequence of the Weierstrass approximation theorem and the uniqueness part
of the Riesz representation theorem that
the moment operators $E^{(k)}$, $k\in\mathbb N$, determine $E$ uniquely.
Since $c_{n,m}=\langle n|E^{(1)}|m\rangle i(n-m)$ for all $n\ne m$,
the phase observable $E$, though not projection valued, is already
determined  by its first moment operator $E^{(1)}$. 
We emphasize that the spectral measure $M$ of the moment operator $E^{(1)}$ is
not phase shift covariant.
Also, the support of $M$ is not necessarily the whole interval $[0,2\pi]$,
and $M$ is {\it never} completely random in the number states.
The latter result follows from the fact that if $M$ was completely random in a number 
state $|n\>$ then the variance must be $\pi^2/3$.
The variance is now
$$
\left\<n\left|\left(E^{(1)}\right)^2\right|n\right\>
-\left(\left\<n\left|E^{(1)}\right|n\right\>\right)^2
=\sum_{m=0 \atop m\ne n}^\infty\frac{|c_{n,m}|^2}{(n-m)^2}
\le\sum_{s=1}^n\frac{1}{s^2}+\sum_{t=1}^\infty\frac{1}{t^2}<\frac{\pi^2}{3}
$$
for all $n\in\mathbb N$. Thus, P1 and P2 (and P0 in many cases) do not hold for $M$.
Note also that for Wigner-Weyl quantized phase angle P1 and P2 do not hold
\cite[p.\ 458]{DuHe}.

The cyclic moments of $E$,
$V^{(k)} = \int_0^{2\pi}e^{ik\theta}\d E(\theta)$, $k\in\mathbb N$,
form another important class of bounded operators constructed from $E$.  They
constitute a nonunitary representation 
$k\mapsto V^{(k)}$
of the additive semigroup $\mathbb N$
provided that $E$ is strong, that is, if $V^{(k)}=(V^{(1)})^k$ for all $k\in\mathbb N$.
This is a necessary  condition for a covariant phase observable to generate number shifts,
that is, for the expression of P3.

\subsection{The canonical covariant phase observable}
The constant matrix $c_{n,m}\equiv1$, 
or, equivalenty, a constant sequence $\psi_n\equiv\psi$ determines
the canonical phase observable
$$
E_{\mathrm{can}}(X):=\sum_{n,m=0}^\infty\frac{1}{2\pi}\int_Xe^{i(n-m)\theta}\mathrm{d}\theta
|n\rangle\langle m|.
$$ 
%$E_{\mathrm{can}}$ is a special case of chess-board phase observables which are 
%determined by the matrix elements $c_{n,m}=1$, when $n-m$ is even, and
%$c_{n,m}=\xi\in[-1,1]$, when $n-m$ is odd.
This is the unique covariant phase observable  associated with
the polar decomposition of the lowering operator
$a=V|a|$, where the partial isometry $V$ is the first cyclic moment $V^{(1)}$ of 
$E_{\mathrm{can}}$ \cite{Holevo}.
{\em The canonical phase $E_{\mathrm{can}}$ is strong, and it 
is (up to unitary equivalence)
the only covariant phase observable which generates number shifts}, that is,
$V^{(k)}|n+k\>=|n\>$ 
for all $k,\,n\in\mathbb N$ \cite{JPPL00}.
The unitary equivalence is to be understood in the sense of covariance systems, which, in the present case
means that two phase observables $E_1$ and $E_2$ are unitarily equivalent if $E_1=
E_2^U:=UE_2U^*$ for some unitary operator $U$ diagonalized 
by the number operator $N$. For example,
$E^U_{\mathrm{can}}$ is a phase observable determined by the phase matrix
$\left(e^{i(\upsilon_n-\upsilon_m)}\right)_{n,m\in\mathbb N}$ where
$\left(\upsilon_n\right)_{n\in\mathbb N}\subset[0,2\pi)$.
We conclude that $E_{\mathrm{can}}$ and $N$ constitute a  true canonical pair:
$N$ generates phase shifts and $E_{\mathrm{can}}$ generates number shifts.

%Also, $\left[N,E_{\mathrm{can}}^{(1)}\right]=iI$ in the dense subspace of $\mathcal H$ \cite{Garrison, Galindo}. 
%In addition, $E_{\mathrm{can}}$ is (up to a unitary equivalence)
%the only covariant phase observable which has a 
%covariant projection valued dilatation on $L^2[0,2\pi)$. The dilatation is the canonical
%spectral measure $X\mapsto\chi_X$ of the multiplication operator
%$(Q\psi)(\theta)=\theta\psi(\theta)$, $\psi\in L^2[0,2\pi)$.

%[The rest of this subsuction is, perhaps, irrelevant for our considerations].

\subsection{Coherent state phase densities and uncertainties}

Let $g^E_{|z\>}$ 
be the probability density of the coherent state phase probability measure $p^E_{|z\>}$,
that is, $p^E_{|z\>}(X)= 
(2\pi)^{-1}\int_Xg^E_z(\theta)\mathrm d\theta.$
Since 
$$
g^E_{|z\>}(\theta)=\sum_{n,m=0}^\infty 
c_{n,m}\frac{|z|^{n+m}}{\sqrt{n!m!}}e^{i(n-m)(\theta-\arg z)},
$$
and $c^{\mathrm{can}}_{n,m}\equiv1$, we have 
$$
g^E_{|z\>}(\arg z)\le g^{E_{\mathrm{can}}}_{|z\>}(\arg z)
$$ 
for any covariant
phase observable $E$ showing that the canonical phase gives the highest value
for the coherent state phase density at $\arg z$. We note also that $g^{E_{\mathrm{can}}}_{|z\>}$ 
%[and also $g^{E_{|s\rangle}}_{|z\>}$] 
tends to a $2\pi$-periodic Dirac $\delta$-distribution in the classical limit 
$|z|\to\infty$ \cite{Garrison, BaPe, JPPL00}.

For $2\pi$-periodic probability densities $g$
the standard deviation is not a good
measure of uncertainty, the appropriate measure being the L\'evy-measure
\cite{Le, Br}
$$
\hbox{L\'evy}(g):=
\inf_{\alpha,\beta\in\mathbb R}\left\{\frac{1}{2\pi}
\int_{\alpha-\pi}^{\alpha+\pi}(\theta-\beta)^2g(\theta)\d\theta\right\}.
$$
For the canonical phase this leads, in coherent states with the amplitude $|z|>1/2$,
to the following phase-number uncertainty relation
$$
\Delta_{|z\rangle} E_{\mathrm{can}}\,
\Delta_{|z\rangle}N\sim\frac{1}{2},
$$
 where
$\Delta_{|z\rangle} E_{\mathrm{can}}=\sqrt{\hbox{L\'evy}\left(g^{E_{\mathrm{can}}}_{|z\>}
\right)}$ and $\Delta_{|z\rangle}N=|z|$ (see \cite{JPPL00, BaPe}).
 
\subsection{Covariant phase space phase observables}
The phase space phase observable generated by a number state $|s\rangle$,
$s\in\mathbb N$, is of the form
$$
E_{|s\rangle}(X):=\frac{1}{\pi}\int_X\int_0^\infty
D\left(re^{i\theta}\right)|s\rangle\langle s|D\left(re^{i\theta}\right)^*\,r\,\mathrm d r\,
\mathrm d\theta,
$$
where $D\left(re^{i\theta}\right) =\exp\left[r\left(e^{i\theta}a^*- e^{-i\theta}a\right)\right]$ 
is the unitary phase space translation operator.
For the ground state $|0\>$ this observable is simply
\begin{eqnarray*}
E_{|0\rangle}(X)&&=\frac{1}{\pi}\int_X\int_0^\infty
\left|re^{i\theta}\right\rangle\left\langle re^{i\theta}\right|\, r\,\mathrm d r\,
\mathrm d\theta\\
&&=
\sum_{n,m=0}^\infty c^{|0\rangle}_{n,m}\frac{1}{2\pi}\int_Xe^{i(n-m)\theta}\mathrm{d}\theta
|n\rangle\langle m|,
\end{eqnarray*}
with the matrix elements
$$
c^{|0\rangle}_{n,m}:=\frac{\Gamma\left((n+m)/2+1\right)}{\sqrt{n!m!}}
$$
(see, e.g.\ \cite{Ari}).
We note that
the angle margin of the Husimi $Q$-function $Q_\varphi(w)=|\<w|\varphi\>|^2, w\in\mathbb C$, 
of a vector state $\varphi\in\mathcal H$, $\parallel\varphi\parallel=1$, is of the form ($w=re^{i\theta}$)
$$
\int_0^\infty Q_\varphi\left(re^{i\theta}\right)\mathrm d r^2
=\left\<\varphi\Big|\frac{\mathrm d E_{|0\rangle}}{\mathrm d\theta}(\theta)\varphi\right\>
=g^{E_{|0\>}}_\varphi(\theta)
$$
where $\frac{\mathrm d E_{|0\rangle}}{\mathrm d\theta}(\theta)=\int_0^\infty
\left|re^{i\theta}\right\>\left\<re^{i\theta}\right|\d r^2$ is an operator density of 
$E_{|0\rangle}$.
Like $g^{E_{\mathrm{can}}}_{|z\>}$, the distribution $g^{E_{|s\rangle}}_{|z\>}$ 
tends to a $2\pi$-periodic Dirac $\delta$-distribution in the classical limit 
$|z|\to\infty$ and 
$\Delta_{|z\rangle} E_{|s\rangle}\,\Delta_{|z\rangle}N\sim\sqrt{\frac{s+1}{2}}$
(see \cite{JPPL00}).
As it stands the only phase observable actually measured is $E_{|0\rangle}$ and it 
can be measured by using e.g.\
the double homodyne detection (see, e.g.\ \cite{Ari}).

\subsection{Positive semidefinite forms}
Fix $J\subseteq\mathbb N$, $J\ne\emptyset$.
Let $\mathcal M_J:=\hbox{lin}\{|n\>\,|\,n\in J\}$, and 
let $\mathcal H_J:=\overline{\mathcal M_J}$, which is a Hilbert subspace of $\mathcal H$.
%be a set of vectors 
%$\sum_{n=0}^\infty s_n|n\rangle\in\mathcal H$ for
%which $\sum_{n=0}^\infty|s_n|<\infty$. 
Let $\mathcal M_J^*$
be the [algebraic] dual of $\mathcal M_J$.
Using the Dirac notation, we thus get $\sum_{n\in J}t_n\langle n|\in\mathcal M_J^*$
for an arbitrary sequence $(t_{n})_{n\in J}\subset\mathbb C$.
Here $\<n|$ denotes the functional $\mathcal H\ni\varphi\mapsto \<n|\varphi\>\in\mathbb C$.
%$\sum_{n=0}^\infty u_n\langle n|\in\mathcal H_1'$ if $(u_{n})_{n\in\mathbb N}\subset\mathbb C$
%is such that $\sup_{n\in\mathbb N}\{|u_n|\}<\infty$. 
%Thus, $\mathcal H_1^{\mathrm{top}}$
%is isomorphic to $\mathcal H_\infty$, and
Recall that
$\mathcal M_J\subseteq\mathcal H_J\simeq\mathcal H_J^*\subseteq\mathcal M_J^*$
and that the equalities hold if and only if the number of elements of $J$
is finite ($\# J<\infty$).

We let $(F|$ denote a generic element of $\mathcal M_J^*$
and we call it a bra vector.
For a given bra vector $(F|\in\mathcal M_J^*$
%resp.  $(F|\in\mathcal H_1'$, 
the ket vector $|F)$ means  the antilinear 
mapping $\mathcal M_J\ni\psi\mapsto\langle\psi|F):=\overline{(F|\psi\rangle}
\in\mathbb C$.
%resp. $\mathcal H_1'\ni\psi\mapsto\langle\psi|F):=\overline{(F|\psi\rangle}\in\mathbb C$. 
For a complex matrix $(d_{n,m})_{n,m\in J}$
the formal double series 
$\sum_{n,m\in J}d_{n,m}|n\rangle\langle m|$ is to be understood as
%can be interpreted to be
the following sesquilinear form:
$$
\mathcal M_J\times\mathcal M_J\ni(\varphi,\psi)\mapsto D(\varphi,\psi):=
\sum_{n,m\in J}d_{n,m}
\langle\varphi|n\rangle\langle m|\psi\rangle\in\mathbb C.
$$
Similarly, we let $|F)(F|$ denote the sesquilinear form 
$
\mathcal M_J\times\mathcal M_J\ni(\varphi,\psi)\mapsto\<\varphi|F)(F|\psi\>\in
\mathbb C.
$

By definition,  matrix $(d_{n,m})_{n,m\in J}$ is  positive semidefinite
if and only if $D(\psi,\psi)\ge0$ for all $\psi\in\mathcal M_J$. We recall that if, in addition,
$J=\mathbb N$ and
$d_{n,n}=1$ for all $n\in\mathbb N$, the matrix $(d_{n,m})_{n,m\in\mathbb N}$ is
a phase matrix.

\begin{theorem}
The following statements  are equivalent:
\begin{itemize}
\item[(i)] $(d_{n,m})_{n,m\in J}$ is  positive semidefinite;
\item[(ii)] $d_{n,m}=\<\psi_n|\psi_m\>$, $n,\,m\in J$, for some vector sequence 
$(\psi_n)_{n\in J}\subset\mathcal H_J$;
\item[(iii)] $d_{n,m}
=\sum_{k\in J}\langle n|F_k)(F_k|m\rangle$,
$(F_k|\in\mathcal M_J^*$, $k\in J$,
and $\sum_{k\in J}|(F_k|n\>|^2<\infty$ for all $n,\,m\in J$.
\end{itemize}
\end{theorem}
\begin{proof}
It is well-known that (i) equals (ii), see, e.g.\  \cite[Chpt 3]{BCR}. 
Let $(\psi_n)_{n\in J}$ be a sequence of vectors in $\mathcal H_J$, and put,
for all $n,\,m\in J$, $d_{n,m}=\<\psi_n|\psi_m\>$. Then
$$
d_{n,m}=\<\psi_n|\psi_m\>= \sum_{k\in J}\<\psi_n|k\>\<k|\psi_m\>=
\sum_{k\in J}\<n|F_k)(F_k|m\>,
$$
where $(F_k|:=\sum_{n\in J}\<k|\psi_n\>\<n|$ and
$\sum_{k\in J}|(F_k|n\>|^2=  \<\psi_n|\psi_n\> <\infty$.

Suppose then that $(F_k|\in\mathcal M_J^*$ 
and $\sum_{k\in J}|(F_k|n\>|^2<\infty$ for all $k,\,n\in J$.
Then, by the Cauchy-Schwarz inequality [for $\ell_{\mathbb C}^2(J)$], the series 
$\sum_{k\in J}\langle\varphi|F_k)(F_k|\psi\rangle$
converges for all $\varphi,\,\psi\in\mathcal M_J$ and it is nonnegative when
$\varphi=\psi$. 
Defining, for all $n,m\in J$,  $d_{n,m}
=\sum_{k\in J}\langle n|F_k)(F_k|m\rangle$,
we see that $D(\varphi,\psi)=\sum_{n,m\in J}d_{n,m}\langle\varphi|n\rangle\langle m|\psi\rangle
= \sum_{k\in J}\langle\varphi|F_k)(F_k|\psi\rangle$ is a positive sesquilinear form, that is,
$(d_{n,m})_{n,m\in J}$ is  positive semidefinite.
\end{proof}

For a positive semidefinite matrix $(c_{n,m})_{n,m\in J}$ with 
$c_{n,n}\equiv1$ one gets
$$
\sum_{n,m\in J} c_{n,m}|n\rangle\langle m|=
\sum_{k\in J}|F_k)(F_k|,
$$
where now $(F_k|\in\mathcal M_J^*$ and $\sup_{k\in\mathbb N}\{|(F_k|n\>|\,:\,n\in J\}\le1$
for all $k\in J$.
%(the notations $\sum_{n,m\in J} c_{n,m}|n\rangle\langle m|$ and $\sum_{k\in J}|F_k)(F_k|$ must be understood as a sesquilinear forms $\mathcal M_J\times\mathcal M_J\to \mathbb C$).
% are the phase states,
%and where the sesquilinear form $\sum_{n,m=0}^\infty c_{n,m}|n\rangle\langle m|$
%can be defined on $\mathcal H_1$.
For the canonical phase, one may choose $\psi_n\equiv|0\rangle$, and thus 
$|F_0)=\sum_{n=0}^\infty|n\rangle$ and $|F_k)=0$ otherwise.
Note that $R(\theta)|F_0)\equiv|\theta):=\sum_{n=0}^\infty e^{in\theta}|n\rangle$ and 
$$
E_{\mathrm{can}}(X)=\frac{1}{2\pi}\int_X|\theta)(\theta|\mathrm d\theta.
$$
For the trivial phase ($c_{n,m}\equiv\delta_{n,m}$) one gets $|F_k)=|k\rangle$,
$\sum_{k=0}^\infty|F_k)(F_k|=I$, and 
$E_{\mathrm{triv}}(X)=(2\pi)^{-1}\int_X\mathrm d\theta\,I$.

\begin{proposition}\label{o}
With the above notations,
$\sum_{n,m\in J} c_{n,m}|n\rangle\langle m|=|F)(F|$ for some 
$(F|\in\mathcal M_J^*$
if and only if $c_{n,m}=e^{i(\upsilon_n-\upsilon_m)}$ for
all $n,m\in J$.
\end{proposition}
\begin{proof}
If $\sum_{n,m\in J} c_{n,m}|n\rangle\langle m|=|F)(F|$, $(F|\in\mathcal M_J^*$, then
$|(F|n\>|^2=1$ for all $n\in J$, and thus $|F)=\sum_{n\in J}e^{i\upsilon_n}|n\>$. Hence
$c_{n,m}=e^{i(\upsilon_n-\upsilon_m)}$ for all $n,m\in J$.
Conversely, if $c_{n,m}=e^{i(\upsilon_n-\upsilon_m)}$ for all $n,m\in J$, then
$\sum_{n,m\in J} c_{n,m}|n\rangle\langle m|=|F)(F|$, with
$|F)=\sum_{n\in J}e^{i\upsilon_n}|n\>$.
\end{proof}
From Proposition \ref{o} one sees that
the canonical phase observable is (up to unitary equivalence)
the only phase observable which can be written in the form
$$
X\mapsto \frac{1}{2\pi}\int_X R(\theta)|F)(F|R(\theta)^*\mathrm d\theta
$$
using a single generalized state $|F)$.

\section{THE PEGG-BARNETT PHASE AND ITS GENERALIZATION}
For a fixed $s\in\mathbb N$, define $\mathcal H_s:=
\mathrm{lin}\{|n\rangle\,|\,n=0,1,...,s\}\subset\mathcal H$, 
and let $\theta_{s,k}:=k\frac{2\pi}{s+1}$, $k=0,1,...,s$.
The Pegg-Barnett (PB) phase states are then  
$$
|\theta_{s,k}\rangle:=\frac{1}{\sqrt{s+1}}\sum_{n=0}^s e^{in\theta_{s,k}}|n\rangle,
$$
and  they determine a spectral measure
\begin{eqnarray}\nonumber
E^{\mathrm{can}}_s\left([a,b)\right)&:=&\sum_{\theta_{s,k}\in[a,b)}|\theta_{s,k}\rangle
\langle\theta_{s,k}|\;\;\;\;\;\;(0\le a<b\le2\pi)\\
&=&\sum_{n,m=0}^s
\underbrace{\frac{1}{s+1}\sum_{\theta_{s,k}\in[a,b)}e^{i(n-m)\theta_{s,k}}}_
{\to\;(2\pi)^{-1}\int_{[a,b)}e^{i(n-m)\theta}\mathrm d\theta}|n\rangle\langle m|.\nonumber
\end{eqnarray}
Since $\|E^{\mathrm{can}}_s\left([a,b)\right)\|\le1$ for all $s\in\mathbb N$ it follows that
\begin{equation}\label{PBlimit}
\hbox{w-lim}_{s\to\infty} E^{\mathrm{can}}_s\left([a,b)\right)=E_{\mathrm{can}}\left([a,b)\right).
\end{equation}
(see also \cite{VaPe}).

Discrete phase shifts $\{\theta_{s,k}\,|\,k=0,1,...,s\}$ with the addition modulo $2\pi$
form a group which is isomorphic to a finite additive group
$\mathbb Z_{s+1}$ [or a cyclic group $C_{s+1}$].
Consider the following unitary representation of 
the group $\mathbb Z_{s+1}$ of discrete phase shifts:
$$
\mathbb Z_{s+1}\ni k+(s+1)\mathbb Z \mapsto R(\theta_{s,k})\in
\mathcal L\left(\mathcal H\right).
$$
This is well defined since $R(\theta_{s,k})=R(\theta_{s,k+(s+1)n})$ for all $n\in\mathbb Z$.
Since $R(\theta_{s,k})|\theta_{s,l}\rangle=|\theta_{s,k+l}\rangle$ one may  easily confirm that
$$
R(\theta_{s,k})E_s^{{\rm can}}(\{\theta_{s,l}\})R(\theta_{s,k})^*=E_s^{{\rm can}}(\{\theta_{s,{k+l}}\})
$$
for all $k,l=0,\ldots,s$, that is, $E_s^{{\rm can}}$ is (discrete) phase shift covariant. We go on to determine
all such POMs.

Let $J\subseteq\mathbb N$, $J\ne\emptyset$, 
$\mathcal H_J=\overline{\mathcal M_J}$, and let $\mathcal L(\mathcal H_J)$
denote the set of bounded operators on $\mathcal H_J$.
Fix an $s\in\mathbb N$, and let  $\mathcal P(D_s)$ denote the power set of
the set $D_s:=\{\theta_{s,k}\in[0,2\pi)\,|\, k=0,1,\ldots,s\}$. 

\begin{theorem}\label{3}
With the above notations, the map
$$
\mathcal P(D_s)\ni X\mapsto E_{J,s}(X) 
:= \sum_{\theta_{s,l}\in X}E_{J,s}(\{\theta_{s,l}\})
\in\mathcal L(\mathcal H_J)
$$
is a normalized positive operator measure
satisfying  the covariance condition
$$
R(\theta_{s,k})E_{J,s}(\{\theta_{s,l}\})R(\theta_{s,k})^*=E_{J,s}(\{\theta_{s,{k+l}}\}),\ 
k,l=0,1,...,s,
$$
 if and only if $E_{J,s}(\{\theta_{s,l}\})$ is of the form
$$
E_{J,s}(\{\theta_{s,l}\})=  %\frac 1{s+1}\sum_{n,m\in J} c^J_{n,m} e^{i(n-m)\theta_{s,l}}|n\>\<m|, \
\frac 1{s+1}R(\theta_{s,l})A^JR(\theta_{s,l})^*,\
k=0,...,s,
$$
where   
%$\sum_{n,m\in J} c^{J}_{n,m}|n\rangle\langle m|$ is a positive operator,
%$c^J_{n,n}=1$ for all $n\in J$, 
$A^J$ is a bounded positive operator on $\mathcal H_J$, with
$\<n|A^J|n\>=1$ for all $n\in J$, 
and,  for any $n\ne m\in J$, if $ \<n|A^J|m\>\ne 0$, then $|n-m|\not\in (s+1)\mathbb Z^+$. 
The covariant POM  $E_{J,s}$ is projection valued if and only if
$\<n|A^J|m\>=e^{i(\upsilon_n-\upsilon_m)}$ for
all $n,\,m\in J$, where $(\upsilon_n)_{n\in J}\subset[0,2\pi)$, and $\# J=s+1$.
\end{theorem}
\begin{proof}
Suppose that $E:\,\mathcal P(D_s)
\to\mathcal L(\mathcal H_J)$ is a POM for which
$R(\theta_{s,k})E_{J,s}(\{\theta_{s,l}\})R(\theta_{s,k})^*=E_{J,s}(\{\theta_{s,{k+l}}\})$.
Choosing $l=0$ one sees that 
$E_{J,s}(\{\theta_{s,k}\})=R(\theta_{s,k})E_{J,s}(\{0\})R(\theta_{s,k})^*$
where $E_{J,s}(\{0\})\in\mathcal L(\mathcal H_J)$ and 
$E_{J,s}(\{0\})\ge O$.

Conversely, if
$E_{J,s}(X)=\sum_{k\in X}R(\theta_{s,k}) B R(\theta_{s,k})^*$, 
$X\in\mathcal P(D_s)$, for some
$B\in\mathcal L(\mathcal H_J)$, $B\ge O$, then $E_{J,s}$ is a covariant positive
operator measure. 
The normalization condition
$$
\sum_{k=0}^sR(\theta_{s,k}) B R(\theta_{s,k})^*=I\big|_{\mathcal H_J}
$$
equals the condition
$$
\<n|B|m\>\underbrace{\sum_{k=0}^s e^{2\pi i(n-m)k/(s+1)}}_
{=(s+1)\delta_{n-m,(s+1)t},\;\;t\in\mathbb Z}=\delta_{n,m},\;\;\;\;n,\,m\in J,
$$
which equals the following two conditions:
\begin{itemize}
\item[(i)] for $n\ne m$, if $\<n|B|m\>\ne0$ then $|n-m|\notin(s+1)\mathbb Z^+$,
\item[(ii)] $\<n|B|n\>=1/(s+1)$ for all $n\in J$.
\end{itemize}
Thus, if $B$ is as above and satisfies conditions (i) and (ii), then define 
$A^J:=(s+1)B$ and the first part of Theorem is proved.

Let $A^J_{n,m}=\<n|A^J|m\>$ for all $n,\,m\in J$.
If $A^J_{n,m}=e^{i(\upsilon_n-\upsilon_m)}$, $n,\,m\in J$, 
where $(\upsilon_n)_{n\in J}\subset[0,2\pi)$ and $\# J=s+1$,
then it is easy to confirm that $E_{s,J}$ is a projection measure.
Conversely, suppose that $E_{s,J}$ is a covariant normalized projection measure, that is,
$$
E_{s,J}(\{\theta_{s,k}\})E_{s,J}(\{\theta_{s,l}\})
=\delta_{k,l}E_{s,J}(\{\theta_{s,k}\})
$$
for all $k,\,l\in\{0,1,...,s\}$.
By direct calculation, one sees that this equals the fact that
\begin{equation}\label{a}
\frac{1}{s+1}\sum_{t\in J}A^J_{n,t}A^J_{t,m}e^{2\pi i[k(n-t)+l(t-m)]/(s+1)}=\delta_{k,l}
A^J_{n,m}e^{2\pi i(n-m)k/(s+1)}
\end{equation}
for all $n,\,m\in J$ and $k,\,l\in\{0,1,...,s\}$.
Multiply both sides of this equation by $e^{-2\pi i k q/(s+1)}/(s+1)$ and 
sum up with respect to $k$ to get
$$
A^J_{n,n-q+u(s+1)}A^J_{n-q+u(s+1),m}=A^J_{n,m},
$$
which holds for all $n,\,m\in J$ and for all
$q,\,u\in\mathbb Z$ for which $n-q+u(s+1)\in J$. Substituting $v:=n-q+u(s+1)$ one gets
\begin{equation}\label{b}
A^J_{n,v}A^J_{v,m}=A^J_{n,m}
\end{equation}
for all $n,\,m,\,v\in J$. Use Eq.\ (\ref{b}) in Eq.\ (\ref{a}) to get
\begin{equation}\label{c}
\sum_{t\in J} e^{2\pi i t(l-k)/(s+1)}=(s+1)\delta_{k,l}.
\end{equation}
From $l=k$ we see that $\# J=s+1$. If $\# J=s+1$ then Eq.\ (\ref{c}) clearly holds.

Since the operator $\sum_{n,m\in J}A^J_{n,m}|n\>\<m|$ is positive
and $A^J_{n,n}\equiv1$ then $|A^J_{n,m}|\le1$ for all $n,\,m\in J$. Using Eq.\
(\ref{a}) with $n=m$ and $k=l$ and the fact that $\# J=s+1$ 
it follows that $|A^J_{n,m}|=1$ for all $n,\,m\in J$.
Due to the positiveness, if $\# J\ge3$ then
$$
\left|
\begin{array}{ccc}
1 &\; A^J_{i,j} \;& A^J_{i,k} \\
&&\\
\overline{A^J_{i,j}} &\; 1 \;& A^J_{j,k} \\
&&\\
\overline{A^J_{i,k}} &\; \overline{A^J_{j,k}} \;& 1
\end{array} \right|\ge0
$$
for all $i,\,j,\,k\in J$ for which $i<j<k$. 
Using this and the condition $|A^J_{n,m}|\equiv1$ one gets by direct calculation
that $A^J_{n,m}=e^{i(\upsilon_n-\upsilon_m)}$, $n,\,m\in J$, 
where $(\upsilon_n)_{n\in J}\subset[0,2\pi)$. If $\# J=2$ or $\# J=1$ this holds 
trivially.

\end{proof}

Since $A^J$ is bounded, we may now write
$$
E_{J,s}(\{\theta_{s,l}\})= \sum_{n,m\in J}\frac 1{s+1} \,A^J_{n,m}\,e^{i(n-m)\theta_{s,l}}\,|n\>\<m|,
$$
where $A^J_{n,m}= \<n|A|m\>$. 
Therefore,  if $\# J=\infty$,
there is no bounded operator $A$ such that $A^J_{n,m}=e^{i(\upsilon_n-\upsilon_m)}$ 
for all $n,m\in J$. In particular, the case $A^J_{n,m}\equiv1$ requires $J$ to be finite.
%If $\# J<\infty$, using the square root operator of  $A=\sum_{n,m\in J}A^J_{n,m}|n\rangle\langle m|$, 
%it follows that $A^J_{n,m}=\langle\psi_n^J|\psi_m^J\rangle$, where $\psi_n^J\in\mathcal H_J$ and $\|\psi_n^J\|=1$ for all $n,\,m\in J$.
%Reordering the terms of $\sum_{n,m\in J} \langle\psi_n^J|\psi_m^J\rangle|n\rangle\langle m|$ one gets $\sum_{n,m\in J} A^J_{n,m}|n\rangle\langle m|=\sum_{k=0}^{\# J-1}|\varphi^J_k\rangle\langle\varphi^J_k|$, where 
%$\varphi^J_k=\sum_n\<\psi_n^J|k\>|n\>\in\mathcal H_J$ for all $k=0,1,...,\# J-1$.

Let $E$ be a covariant phase observable with the phase matrix 
$(c_{n,m})_{n,m\in\mathbb N}$.
Let $J\equiv J(s):=\{0,1,...,s\}$.
If $\lim_{s\to\infty}A^{J(s)}_{n,m}=c_{n,m}$ for all $n,\,m\in\mathbb N$ then
$\hbox{w-lim}_{s\to\infty} E_{J(s),s}\left([a,b)\right)=E\left([a,b)\right)$.
The sequence  $(E_{J(s),s})_{s\in\mathbb N}$ gives an approximation sequence for $E$.
In this way we have generalized the PB-formalism to cover all covariant phase 
observables. Indeed, for a given phase matrix $(c_{n,m})_{n,m\in\mathbb N}$
one may choose, for example, $A^{J(s)}_{n,m}=c_{n,m}$ for $n,\,m\le s$ to generalize
(\ref{PBlimit}).
Especially, the operators $E_{|0\rangle}\left([a,b)\right)$ 
of the phase observable $E_{|0\rangle}$ 
can be approximated weakly by the sequence of operators
$$
\sum_{n,m=0}^s \frac{\Gamma\left((n+m)/2+1\right)}{\sqrt{n!m!}}\,
\frac{1}{s+1}\sum_{\theta_{s,k}\in[a,b)}e^{i(n-m)\theta_{s,k}}
|n\rangle\langle m|.
$$
when $s\to\infty$.

The approximation sequence  $(E_{J(s),s})_{s\in\mathbb N}$ of
a phase observable $E$ gives a discretization of $E$ when
$\hbox{w-lim}_{s\to\infty} E_{J(s),s}\left([a,b)\right)=E\left([a,b)\right)$,
and it satisfies the spectral accuracy condition \cite[p.\ 495]{DuHe}
$$
\sup_{\theta\in[a,b)}\left(\min_{\lambda\in D_s}|\theta-\lambda|\right)\to0
$$
for all $a<b\in[0,2\pi]$
when $s\to\infty$.

Theorem \ref{3} shows that only $E^U_{\mathrm{can}}$
has a projection valued discretization.
In this sense one may say that $E^U_{\mathrm{can}}$ is almost projection valued.
This is somewhat striking since $E^U_{\mathrm{can}}$ is known
to be totally noncommutative \cite{vimppa}.
Also, from Proposition \ref{o} one obtains that $E_{\mathrm{can}}$ is (up to unitary equivalence) the only 
phase observable which is determined by only one (generalized) phase state $|\theta)$,
and it is (up to unitary equivalence) the only phase observable which has the
approximation sequence determined by one discrete phase state $|\theta_{s,k}\>$.
That is, $E^U_{\mathrm{can}}$ can be approximated by the operator measures
of the form
$$
\frac{1}{s+1}\sum_{\theta_{s,k}\in[a,b)}R(\theta_{s,k})|F)(F|R(\theta_{s,k})^*.
$$
\newline

\noindent
{\bf Acknowledgments.}
We are grateful to Stephen Barnett 
for fruitful discussions during his visit in Turku, 
concerning the connections between the PB-formalism and
the covariant approach. 
In particular,
his assistance in the development of the proof of Theorem \ref{3}
is warmly acknowledged. 
We also thank Paul Busch for his 
suggestions to improve the final form of the article.

\end{document}